\documentclass{article}

\usepackage{hyperref}
\usepackage[utf8]{inputenc}
\usepackage{latexsym,graphicx,amssymb,amsmath} 

\newcommand{\bea}{\begin{eqnarray}}
\newcommand{\eea}{\end{eqnarray}}
\newcommand{\be}{\begin{equation}}
\newcommand{\ee}{\end{equation}}

\newcommand{\rt}[1]{{}}
\newlength{\szovszel}
\newlength{\slashszel}

\begin{document}

\title{Bound-state spectra of field theories\\ through separation of external and internal dynamics}

\author{A. Jakov\'ac and A. Patk\'os\\
Institute of Physics, E\"otv\"os University,\\
Budapest, H-1117, Hungary\\
E-mail: antal.jakovac@gmail.com,patkos@hector.elte.hu}

\maketitle

\begin{abstract}
A general strategy is formulated for computing bound state spectra in the framework of functional renormalisation group (FRG). Dynamical "coordinates" characterising bound states are introduced as coupling parameters in the $n$-point functions of effective fields representing the bound states in an extended effective action functional. Their scale dependence is  computed with functional renormalisation group equations.
In the infrared an interaction potential among the constituting fields is extracted as smooth function of the coupling parameters. Eventually quantised bound state solutions are found by solving the Schr\"odinger eigenvalue problem formulated for the coupling parameters transmuted into coordinates. The proposed strategy is exemplified through the analysis of a recently published FRG study of the one-flavor chiral Nambu--Jona-Lasinio model.
{\footnote[1]{Invited paper to appear in Gribov-90 Memorial Volume, Editors: Yu. Dokshitzer, P. L\'evai, \'A. Luk\'acs and J. Nyiri, World Scientific 2020}}
 
\end{abstract}

\section{Introduction and motivation}

Bound states of two particles appear as (complex) pole singularities in two-particle propagators. In practice this propagator is parametrised with finite number of intuitively chosen parameters. Approximate solution of the bound state problem consists of finding optimised values of the parameters, reflecting the expected qualitative physical features.

A widely used procedure in quantum chemistry is the Born-Oppenheimer approximation\cite{BO1927}, where the electronic wave function is parametrised with a fixed static distance of the two nuclei. The corresponding Schr\"odinger energy eigenvalues are smooth functions of the nuclear coordinates. The optimisation step consists of the determination of the probability amplitude of the distance distribution, which is realised by solving the Schr\"odinger-equation for the quantum motion of the nuclei. Similar approach is used in heavy quark spectroscopy of QCD, where in a first step the interquark potential is determined through the exchange of dynamical gluons and quarks between static colored sources, and next the Schr\"odinger-problem of the quark sources is solved in this potential\cite{HQ-spectr}. The interquark potential is found by measuring the correlator of two Polyakov-lines on a space time lattice, and extracting the renormalised interaction potential after appropriately subtracting the self-energies of the individual lines\cite{Polyakov-renorm}. In both physical problems the binding energy (the missing mass) is orders of magnitude smaller than the complete mass of the composite. 

The quality of these approximate scenarios depends critically on the decoupling of the dynamics of the "force field sources" from the rest of dynamical degrees of freedom and also on the hierarchical ordering the subsequent contributions to the complete energy. It is highly desirable to construct a systematic procedure which can test whether the internal dynamics of the candidate subsystems is influenced only through some collective effects (like the interaction potential) emerging from the motion of the rest of the complete system. One possible approach is to introduce collective fields/wave functions representing the prospective bound state with help of an auxiliary function. 

In this short note a strategy is put forward to determine a smooth interrelation between different parameters characterising the propagator and the coupling of the auxiliary field representing the composite (bound) degree of freedom to the original fields. Functional renormalisation group equations are ideal in searching for these functional relations. On the basis of such stable, physically meaningful relations one can proceed to the second stage and solve the quantum equations for the reduced set of parameters describing the internal quantum dynamics of the composite field.   
The proposition for a general strategy is described in detail in section 2. It is applied in section 3 to the symmetric phase of the chiral Nambu--Jona-Lasino model for which an interaction potential has been determined non-perturbatively among the fermionic constituents recently\cite{jakovac20}.

This specific model, where chiral symmetry forces the defining fields massless, represents some additional interest to us. The very accurate {\it ab initio} reconstruction of the lowest lying baryon spectra with light quarks\cite{fodor08} is one of the greatest success of lattice field theory. However, the lattice approach does not offer any insight on the emergence of the concept of constituent quark mass, which is the basis of the widest used non-relativistic quark models. Effective models  of chiral dynamics from the earliest days\cite{chiral-eff-models} relate this mass to the chiral condensate of strong interactions, which raises, however, the question of the existence of hadronic bound states in the phase of restored chiral symmetry. In a broader context of the Standard Model the mechanism of producing vector and scalar bound states with light (relative to the Planck mass) fermions was a central problem also for V.N. Gribov\cite{gribov-bonn-95}.

\section{The strategy}

Consider a quantum theory of defining fields $\varphi(x)$. Its solution is encoded into the effective action $\Gamma[\varphi]$ from which all $n$-point functions can be extracted through appropriate functional derivatives. The straightest way to look for bound states formed with $N=2,3,...$ constituent fields is to look into the analytic structure of the $2N$-point functions. This program is technically difficult, maybe even impossible to realize. It is more realistic to consider a collective field introduced in the channel where one searches for the existence of a bound state:
\begin{equation}
H(x)\leftrightarrow\int[\Pi_{i=1}^{N} dy_i]O(x-y_1,...,x-y_N)\varphi(y_1)...\varphi(x_N)
\label{multifield-replacement}
\end{equation}
where $O(z_1,..z_N)$ characterizes the space-time stucture of the compound system. The collective field is introduced into the theory as an auxiliary field, for instance with help of the quadratic expression:
\begin{eqnarray}
&\displaystyle
\Delta\Gamma[H,\varphi]=\nonumber\\
&\displaystyle
\frac{M_H^2}{2}\int dx\left[H(x)-\frac{g}{M_H^2}\int[\Pi_{i=1}^{N} dy_i]O(x-y_1,...,x-y_N)\varphi(y_1)...\varphi(x_N)\right]^2,
\label{delta-eff-action}
\end{eqnarray}  
with scale dependent new couplings $M_H^2,g$.
The hunting for the bound state focuses now on the two-point function of $H(x)$ taking into account the effect of the quantum fluctuations
of the field $\varphi$ by running the renormalisation group equations formulated for the Euclidean theory \cite{wetterich93,morris94}. Theories extended this way were used for investigating bound states in Refs.\cite{ellwanger94,gies02,pawlowski07} (see also the recent careful analysis of Ref.\cite{jakovac19}). Detailed discussion of the mesonic bound state spectra was based on this kind of transformed QCD first in Ref.\cite{jungnickel96}. 
In this investigation composite fields were defined locally, without any internal structure. As a consequence only the first stage of the strategy to be outlined below has been realized and the renormalized mass-spectra without including any effect of the internal dynamics of the constituents was fitted to the observed meson spectra.
 
The trial two-point function is parametrised in the Euclidean version of the theory in a way reflecting the expected occurrence of a pole:
\begin{equation}
G_H(p)=\frac{Z_H}{p^2+M_H^2}+{\textrm{polynomial background}}.
\end{equation}
The infrared values of the scale dependent parameters are controlled by the evolution of the coupling (vertex) function $O(z_1,...,z_N)$ weighting the contributions emerging from the interacting defining fields. For instance the simple trial form for the Fourier transformed vertex function
\begin{equation}
O(q_1,q_2,...,q_N)\sim \Pi_{i\neq j} e^{-\alpha_{ij} (q_i-q_j)^2}\times e^{-\beta(Q-\sum q_i)^2}
\label{vertex-parametrisation}
\end{equation}
introduces a spatio-temporal range $\alpha=\sup_{ij}\{\alpha_{ij},\beta\}$ to which all constituents are restricted. One can easily invent higher cluster distance restrictions. 

The interesting case is when the scale dependent parameters stay close to their classical (UV) values and their infrared values change smoothly (slowly) with the input (initial) values. Under this assumed behavior one can anticipate the existence of a smooth functional relation between the important structural parameters of the effective action, for instance
\begin{equation}
M_H^2(IR)=M_H^2(\alpha(IR)).
\end{equation}
This function is the central object of the proposed strategy. Its existence cannot be guaranteed, it is expected only intuitively. 
The next task is to deconstruct carefully this function. For large $\alpha_0(UV)$ one expects the particles corresponding to $N$ fields to fill uniformly the space of linear size $\alpha_0^{1/2}$. One has to expect that $M_H^2(IR,0)$ tends to a limiting value when $\alpha_0$ increases beyond any limit. For this limiting case the change $\delta M_H^2(0)\equiv M_H^2(IR,0)-M_H^2(UV,0)$ can be interpreted as the sum of one-particle self-energy contibutions to the invariant squared "mass", since the initial value $M_H^2(UV)$ is the classical squared mass of the $N$-field complex. The interesting question concerns what happens with this difference when $\alpha$ (e.g. the available volume) is diminished gradually?

Generically, for $\alpha_1(UV)<\alpha_0(UV)$ one finds $\delta M_H^2(1)\equiv M_H^2(IR,1)-M_H^2(UV,1)\neq \delta M_H^2(0)$. The difference reflects the interaction among the constituents. One can map out the dependence of this interaction squared energy on the squared size by gradually changing $\alpha(UV)$. This simple disentanglement of the different contributions can be expressed formally for any given $\alpha$ as
\begin{equation}
M_H^2(IR) = M_H^2(UV)+\delta M_H^2(0) +\Delta M_H^2({\textrm{interaction}}).
\label{energy-disentanglement}
\end{equation}
 
The second part of the strategy consists of solving the quantum mechanical N-body problem with identical particles of squared classical mass  $m_H^2=(M_H^2(UV)+\delta M_H^2(0))/N$ moving in the generalized potential $[\Delta M_H^2(\{\alpha_{ij}\})]^{1/2}$. $m_H$ corresponds to the constituent mass formed dynamically. If one uses the relative coordinates $(\alpha_{i<j})^{1/2}$ the simplest Hamiltonian defining the quantum mechanical problem is
\begin{equation}
\hat H=\sum_{i<j}\frac{\hat \pi_{ij}^2}{2m_H}+[\Delta M_H^2({\textrm{interaction}})]^{1/2},
\label{reduced-Hamiltonian}
\end{equation}
where $\hat\pi_{ij}$ is the momentum conjugate to $\alpha_{ij}^{1/2}$.
Bound state solutions of this system correspond to eigenvalues lower than $(M_H^2(UV)+\delta M_H^2(0))^{1/2}$. The difference is the binding energy. By the examples quoted in the introductory part one expects good quality results if $m_H>>[\Delta M_H^2({\textrm{interaction}})]^{1/2}$.

In conclusion, this general strategy based on extracting the renormalised interaction potential with help of renormalisation group equations in principle offers an equivalent procedure to the numerical simulation and renormalisation of the appropriate correlations in the framework lattice field theories. In the next section a concrete realisation of the above general strategy is presented.

\section{Two-particle bound state in the symmetric phase of the one-flavor chiral NJL model}

\begin{figure}[htbp]
\begin{center}
\includegraphics[width=2in]{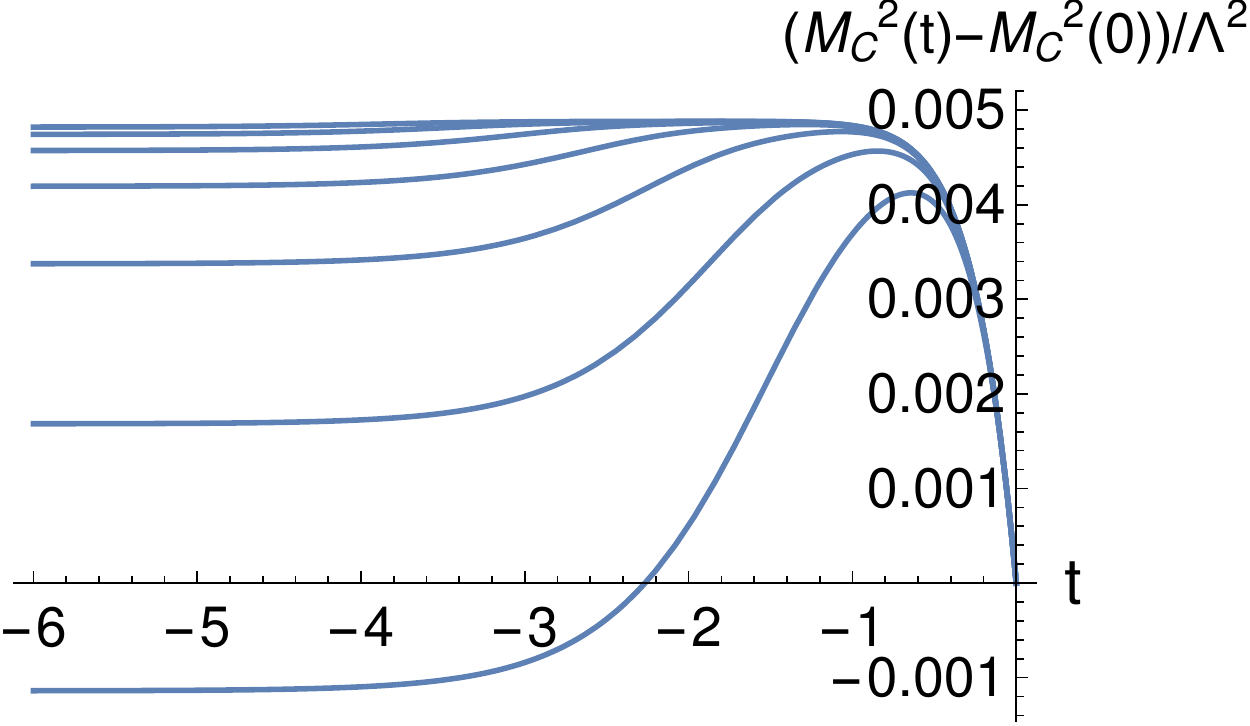}
\end{center}
\caption{RG-variation (as a function of $t=\ln(k/\Lambda)$) of the quantum contribution to the squared composite mass $\delta M_C^2(\alpha(k=\Lambda))=M_C^2(t=-\infty,\alpha(k=\Lambda))-M_C^2(t=0,\alpha(k=\Lambda))$ in units of $\Lambda^2$. From the top to the bottom curves with diminishing $\alpha_r(k=\Lambda)$ are presented. From Ref.\cite{jakovac20}}
\label{alpha_mphys2_vs_t}
\end{figure}

In our recent paper we have introduced a collective field in the sense of (\ref{multifield-replacement}) for the degenerate scalar-pseudoscalar two-fermion sector of the chiral NJL-model. A single "slow" variable $\alpha(UV)$ was introduced in a Gaussian ansatz like   (\ref{vertex-parametrisation}). Infrared values of $M_S^2=M_{PS}^2\equiv M_C^2(t=-\infty)$ were obtained by solving a coupled set of Wetterich equations. Slowly tuning $\alpha(UV)$ from 350 to 2.85 smooth variation of $M_C^2(IR,\alpha(UV))$ was detected (see Fig.\ref{alpha_mphys2_vs_t}).

Next we have performed the analysis summarized in Eq.(\ref{energy-disentanglement}) resulting in the smooth 
 $\Delta M^2_C(\alpha(UV)\Lambda^2)/\Lambda^2$ function (see Fig.\ref{int-energy-size}):
\begin{figure}[htbp]
\begin{center}
\includegraphics[width=2.4in]{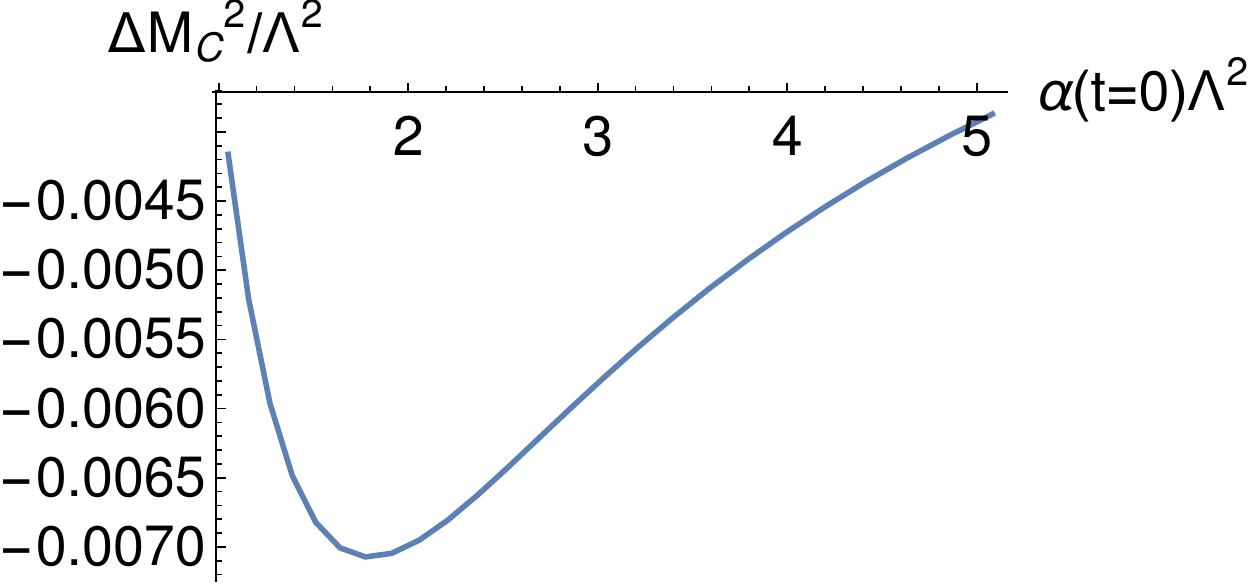}
\end{center}
\caption{Dependence of the interaction energy of the composite on $\alpha(t=0)\Lambda^2=\alpha(k=\Lambda)\Lambda^2$ the cut-off value of the size parameter. From Ref.\cite{jakovac20} }
\label{int-energy-size}
\end{figure}
From this figure one can extract the dimensionless curvature $d^2[\Delta M_C^2]^{1/2}/ d(\alpha^{1/2})^2/\Lambda^3$ at the minimum to be denoted (for dimensional reasons) by $\Omega^3$. To a very good approximation the dimensionless reduced mass of the two-particle state is $\mu_C\equiv [M_C^2]^{1/2}/2\Lambda$. The Schr\"odinger eigenvalue equation for the dimensionless binding energy $\epsilon=E/\Lambda$ can be written with help of the dimensionless radial momentum $p=\pi/\Lambda$ and the dimensionless radial distance $x=\alpha^{1/2}\Lambda$ as 
\begin{equation}
\left(\frac{1}{2}\frac{p^2}{\mu_C}+\frac{1}{2}\Omega^3x^2\right)\Psi=\epsilon \Psi.
\end{equation}
It is important to note that near the minimum $(\alpha(t=0)\Lambda^2)^{-1}\sim 1/2$ which is much larger than $|\Delta M_C^2/\Lambda^2|\sim 0.007$, therefore one consistently can use non-relativistic quantum theory for dealing with the internal dynamics.
 
\begin{figure}[htbp]
\begin{center}
\includegraphics[width=2in]{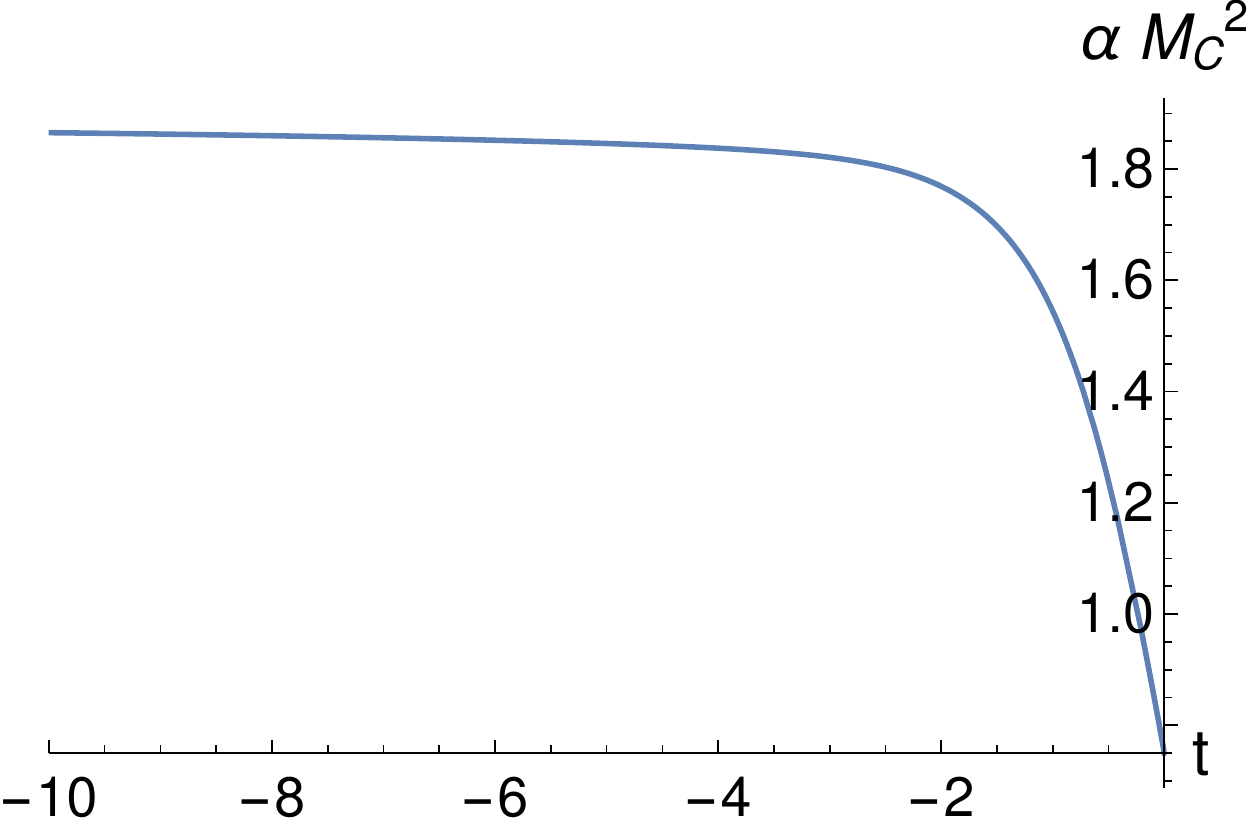}
\end{center}
\caption{RG-variation of the product of the squared boson mass and of the width of the composite "wave function". From Ref.\cite{jakovac20}}
\label{MC-alpha-v}
\end{figure}

One can estimate the ground state energy using the uncertainty principle of Heisenberg in the form $px\sim 1$. An important peculiarity of the present NJL system is the observation made upon Fig.\ref{MC-alpha-v}, namely in the infrared limit one has $x\mu_C\rightarrow {\textrm{const.}}\equiv K$.
The limiting value of the constant is reached very slowly and it very weakly depends on $\alpha(UV)$. It equals approximately 2. The best approximation is to take $ K=x_{min}\mu_C(x_{min})$.
 
This leads to
\begin{equation}
\hat H=\Lambda\left(\frac{x}{2K}p^2+\frac{1}{2}\Omega^3x^2+\epsilon_{min}\right),
\end{equation}
where $\epsilon_{min}\approx -0.007$ by the figure.
Replacing $x$ everywhere by $1/p$ in view of the uncertainty relation one finds the condition for the extremum of $\epsilon(p)$:
\begin{equation}
\frac{d\epsilon}{dp}=\frac{1}{2K}-\frac{1}{p^3}\Omega^3=0
\end{equation}
which gives for the ground state energy
\begin{equation}
\epsilon=\frac{3}{2}(2K)^{-2/3}\Omega+\epsilon_{min}.
\end{equation}

\section{Conclusions}

In this note we proposed a strategy for extracting the renormalised interaction potential of the constituting objects of a bound state from the renormalised squared mass parameter $M_H^2$ defined in (\ref{delta-eff-action}). This potential might turn out a smooth continuous function of the length-like parameters $\sqrt{\alpha_{ij}}$ characterising the function $O(x-y_1,...,x-y_N)$ linking the constituents to the composite field $H(x)$ representing the bound state. The renormalisation group evolution of the parameters $M_H^2,\alpha_{ij}$ as described, for instance, by the Wetterich equation, is the result of the action of the fluctuations of the elementary fields defining the model. The internal quantum dynamics of the constituents of the bound state leads to discrete energy levels. This dynamics is defined in an admittedly intuitive step, adding non-relativistic kinetic terms to the potential energy defined through momentum variables canonically conjugate to the length-like parameters and a constituent mass emerging from the infrared limit of $M_H^2$. Of course the consistency of the non-relativistic nature of the dynamics should be checked.

\section*{Acknowledgements}
This research was supported by the Hungarian Research Fund under the contract K104292. The authors are indebted to dr. J\'ulia Nyiri for the invitation to contribute to the Gribov-90 Memorial Volume.


\begin{thebibliography}{0}
\bibitem{BO1927}M. Born and R. Oppenheimer, Annalen d. Physik  {\bf 398} (1927)
\bibitem{HQ-spectr}  J.L. Richardson, Phys.Lett. B{\bf 82} (1979) 272-274
\bibitem{Polyakov-renorm} O. Kaczmarek, F. Karsch, P. Petreczky,  and F. Zantow, Phys. Lett. B{\bf 543} (2002) 41
\bibitem{jakovac20}A. Jakov\'ac and A. Patk\'os, Mod. Phys. Lett. A{\bf 35} (2020) 2050130
\bibitem{fodor08}S. Durr, Z. Fodor, J. Frison, C. Hoelbling, R. Hoffmann, S.D. Katz, S. Krieg, T. Kurth, L. Lellouch, T. Lippert, K.K. Szabo, G. Vulvert, Science {\bf 322} (2008) 1224
\bibitem{chiral-eff-models}Y. Nambu and G. Jona-Lasinio, Phys. Rev. {\bf 122} (1961) 345, {\it ibid.} {\bf 124} (1961) 246
\bibitem{gribov-bonn-95}V.N. Gribov, {\it Bound States of Massless Fermions as a Source for New Physics}, Bonn TK-95-35, published in: V.N. Gribov, Gauge Theories and Quark Confinement, PHASIS, Moscow, 2002, pp.483-496
\bibitem{wetterich93}C. Wetterich, Phys. Lett. B{\bf 301} (1993) 90
\bibitem{morris94}T.R. Morris, Int. J. Mod. Phys. A{\bf 6} (1994) 2411
\bibitem{ellwanger94}U. Ellwanger and C. Wetterich, Nucl. Phys. B{\bf 423} (1994) 137
\bibitem{gies02}H. Gies and C. Wetterich, Phys. Rev. D{\bf 65} (2002) 065001 
\bibitem{pawlowski07} J.M. Pawlowski, Ann. Phys. (N.Y.) {\bf 322} (2007) 2831
\bibitem{jakovac19} A. Jakov\'ac and A. Patk\'os, Int. J. Mod. Phys. A{\bf 34} (2019) 1950154
\bibitem{jungnickel96} D.U. Jungnickel and C. Wetterich, Phys. Rev. D{\bf 53} (1996) 5142 
\end{thebibliography}
\end{document}